\documentclass[aps,prx,twocolumn,citeautoscript,footinbib,eqsecnum]{revtex4-2}
\usepackage{feynmp}

\usepackage{subcaption}
\usepackage{amsmath}
\usepackage{amssymb}
\usepackage{bbm}
\usepackage{braket}
\usepackage{xcolor}
\usepackage{pifont}
\usepackage{slashed}
\usepackage[mathscr]{euscript}
\usepackage[shortlabels]{enumitem}
\usepackage{tikz-feynman}
\usepackage[papersize={8.5in,11in}]{geometry}

\allowdisplaybreaks

\usepackage{graphicx}
\usepackage{subcaption}
\usepackage[colorlinks=true]{hyperref}
\usepackage{comment}
\usepackage{multirow}

\renewcommand{\approx}{\simeq}
\renewcommand{\Re}{\text{Re}}

\renewcommand{\vec}[1]{\boldsymbol{#1}}
\newcommand{\vn}[1]{|\boldsymbol{#1}|}

\definecolor{wrongultramarine}{rgb}{1,0.5,0}

\newcommand{\rd}{{\rm d}}
\newcommand{\sgn}{{\rm sgn\,}}

\newcommand{\calN}{{\mathcal N}}

\newcommand{\calH}{{\mathcal H}}
\newcommand{\calO}{{\mathcal O}}
\newcommand{\verteq}{\rotatebox{90}{$\,=$}}
\newcommand{\vertdots}{\rotatebox{90}{$\,\dots$}}

\hypersetup{
    bookmarks=true,         
    unicode=false,          
    pdftoolbar=true,        
    pdfmenubar=true,        
    pdffitwindow=false,     
    pdfstartview={FitH},    
    pdfsubject={},   
    pdfcreator={},   
    pdfproducer={}, 
    pdfkeywords={} {} {}, 
    pdfnewwindow=true,      
    colorlinks=true,       
    linkcolor=blue, 
    citecolor=blue,        
    filecolor=magenta,      
    urlcolor=blue           
}

\tikzset{
  mid arrow/.style={postaction={decorate,decoration={
        markings,
        mark=at position .575 with {\arrow[#1]{stealth}}
      }}},
  near arrow/.style={postaction={decorate,decoration={
        markings,
        mark=at position .275 with {\arrow[#1]{stealth}}
      }}},
   far arrow/.style={postaction={decorate,decoration={
        markings,
        mark=at position .800 with {\arrow[#1]{stealth}}
      }}},
   boson/.style={decorate, draw=black,
    decoration={snake,amplitude=1pt, segment length=5pt},
      },
   mid triangle/.style={postaction={decorate,decoration={
        markings,
        mark=at position .575 with {\arrow[#1]{triangle 45}}
      }}}
}

\begin{document}

\title{Fluctuation Spectrum of 2+1D Critical Fermi Surface and its Application to Optical Conductivity and Hydrodynamics}
\begin{abstract}
    We extend the kinetic operator formalism developed in the companion paper [\href{https://doi.org/10.48550/arXiv.2311.03455}{H.Guo,arXiv:2311.03455}] to study the general eigenvalues of the  fluctuation normal modes. We apply the formalism to calculate the optical conductivity of a critical Fermi surface near the Ising-Nematic quantum critical point. We find that the conductivity is the sum of multiple conduction channels including both the soft and non-soft eigenvectors of the kinetic operator, and therefore it is not appropriate to interpret the optical conductivity using extended Drude formula for momentum conserved systems. We also show that the propagation of the FS soft modes is governed by a Boltzmann equation from which hydrodynamics emerges. We calculate the viscosity and it shows clear signature of the non-Fermi liquid physics.
\end{abstract}

\author{Haoyu Guo}
\affiliation{Laboratory of Atomic and Solid State Physics, Cornell University,
142 Sciences Drive, Ithaca NY 14853-2501, USA}

\date{\today}

\maketitle
\tableofcontents

\section{Introduction}

    The critical Fermi surface model of a Fermi surface (FS) coupled to a gapless bosonic field is a toy model for various finite-density quantum matter \cite{SSLee2018,PALee1989,AJMillis1993,JPolchinski1994, BIHalperin1993,YBKim1994a,CNayak1994,SSLee2009,MAMetlitski2010,
    DFMross2010,SSur2014,MAMetlitski2015,SAHartnoll2014,
    AEberlein2017,THolder2015,THolder2015a,ALFitzpatrick2014,
    JADamia2019,JADamia2020,JADamia2021,SPRidgway2015,AAPatel2018b,
    DChowdhury2018c,EGMoon2010,AAbanov2020,YMWu2020,AVChubukov2020,
    XWang2019,AKlein2020,OGrossman2021,DChowdhury2020,IEsterlis2019,
    DHauck2020,YWang2020a,EEAldape2022,AAPatel2017a,AAPatel2019,
    VOganesyan2001,AAPatel2018,AVChubukov2017,DLMaslov2017,
    SLi2023,Iesterlis2021,HGuo2022a,HGuo2023a,LVDelacretaz2022a,
    SEHan2023,UMehta2023,DVElse2021a,DVElse2021,ZDShi2022,ZDShi2023,TPark2023}. In particular, this model is believed to describe the non-Fermi liquid (NFL) state, where the strong fermion-boson coupling destroys the coherent fermionic quasiparticles. Starting from the seminal work by Sung-Sik Lee \cite{SSLee2009}, various large-$N$ approaches have been proposed to analytically control the problem including dimensional/codimensional regularizations of the FS or deforming the boson dispersion (see \cite{SSLee2018} for review), or theories inspired by the Sachdev-Ye-Kitaev model that utilized random coupling in flavor space \cite{DChowdhury2022,Iesterlis2021,HGuo2022a,AAPatel2023,HGuo2023a,ZDShi2022,ZDShi2023}. While these large-$N$ theories differ in detail, their leading order behavior conforms to the Migdal-Eliashberg framework \cite{FMarsiglio2020}, where the vertex correction of the Yukawa coupling is ignored in the computation of the self-energies of the fermion and the boson.

    In this paper we study the transport properties of a translational invariant critical Fermi surface in 2+1 spacetime dimensions. The uniform conductivity $\sigma(\omega)$ is heavily constrained by momentum conservation. First, because of the nonzero overlap between the current and the momentum, the DC conductivity $\sigma_\text{DC}$ diverges (we ignore umklapp effects). Second, the finite-frequency optical conductivity $\sigma(\omega)$ is constrained by statements like Kohn's theorem \cite{WKohn1961}. In particular, the proposed $|\omega|^{-2/3}$ correction \cite{YBKim1994a} to the Drude conductivity is cancelled for a Galilean invariant band structure \cite{DLMaslov2011,HKPal2012,HGuo2022a,ZDShi2022} and the scaling of the next order term is not settled. Another issue is the appropriateness of the extended Drude model \cite{PBAllen1971,JWAllen1977,NPArmitage2018} for optical conductivity. While it is always mathematically correct to parameterize $\sigma(\omega)$ with a frequency-dependent effective mass $m(\omega)/m$ and scattering rate $1/\tau(\omega)$, the physical picture implied in this parameterization is that the dissipations are related to the Drude peak, which is related to the physics of momentum dissipation. This interpretation is not compatible with a translational invariant system where momentum is conserved.

    Part of the goals of this paper is to provide an answer for the questions raised above. We utilize the formalism developed in the companion paper \cite{prlpaper}, where we obtained the fluctuation spectrum of a 2+1D critical FS by diagonalizing the Bethe-Salpeter kernel $K_\text{BS}$ within the Migdal-Eliashberg theory. The optical conductivity can then be evaluated in a spectral decomposition fashion by projecting the current operator onto eigenmodes of $K_\text{BS}$. The optical conductivity we obtained can be written as the sum of various conduction channels and only one of them is the Drude peak, in contrast with the extended Drude model. When the band structure is Galilean invariant (i.e. parabolic), only the Drude peak contribution survives and we recover $\sigma(\omega)=ne^2/(-i\omega m)$. When the band is not Galilean invariant, there will be contributions from other channels and also corrections to the Drude peak. The conduction channels can be roughly classified into two types. The first type has nonzero overlap with the current at zero energy, which includes the momentum channel yielding the Drude peak and odd-parity deformations of the FS which has a slow relaxation rate \cite{prlpaper}. The second type are operators that only overlap with the current at finite energy and they produce an incoherent contribution to the conductivity.

    While the uniform conductivity is dominated by the Drude peak where the critical FS physics only enters as a correction, the non-local conductivity $\sigma(\vec{p})$ is a more informing quantity that reflects the critical FS physics. This is because momentum conservation facilitates the emergence of hydrodynamic transport  \cite{RNGurzhi1968,RJaggi1991,MJMdeJong1995,MMueller2008,MMuller2009,AVAndreev2011,ALucas2016a,BNNarozhny2015,APrincipi2016,LLevitov2016,GFalkovich2017,DABandurin2016,
    JCrossno2016,LVDelacretaz2019,HGuo2017,HGuo2017a,HGuo2018,RKrishnaKumar2017}. Another goal of this paper is to provide a description of the hydrodynamic transport of the critical FS. At the longest length scale, the fermion-boson fluid is described by the Navier-Stokes equation with a viscosity term $\nu$ and we demonstrate that this viscosity is directly related to the soft eigenvalue $\lambda_2$ of $K_\text{BS}$ which is derived in the companion paper \cite{prlpaper}. Because $K_\text{BS}$ contains a large number of soft modes, at an intermediate length scale $L<L_\text{tomo}$ all these soft modes propagate collectively in space and the system enters the tomographic transport  regime \cite{PLedwith2019,SKryhin2023}. This regime is characterized by a scale-dependent viscosity $\nu(\vec{p})$ and its scaling property is a non-perturbative synthesis of the soft mode eigenvalues $\lambda_m$.

\begin{figure}
  \centering
  \includegraphics[width=1.0\columnwidth]{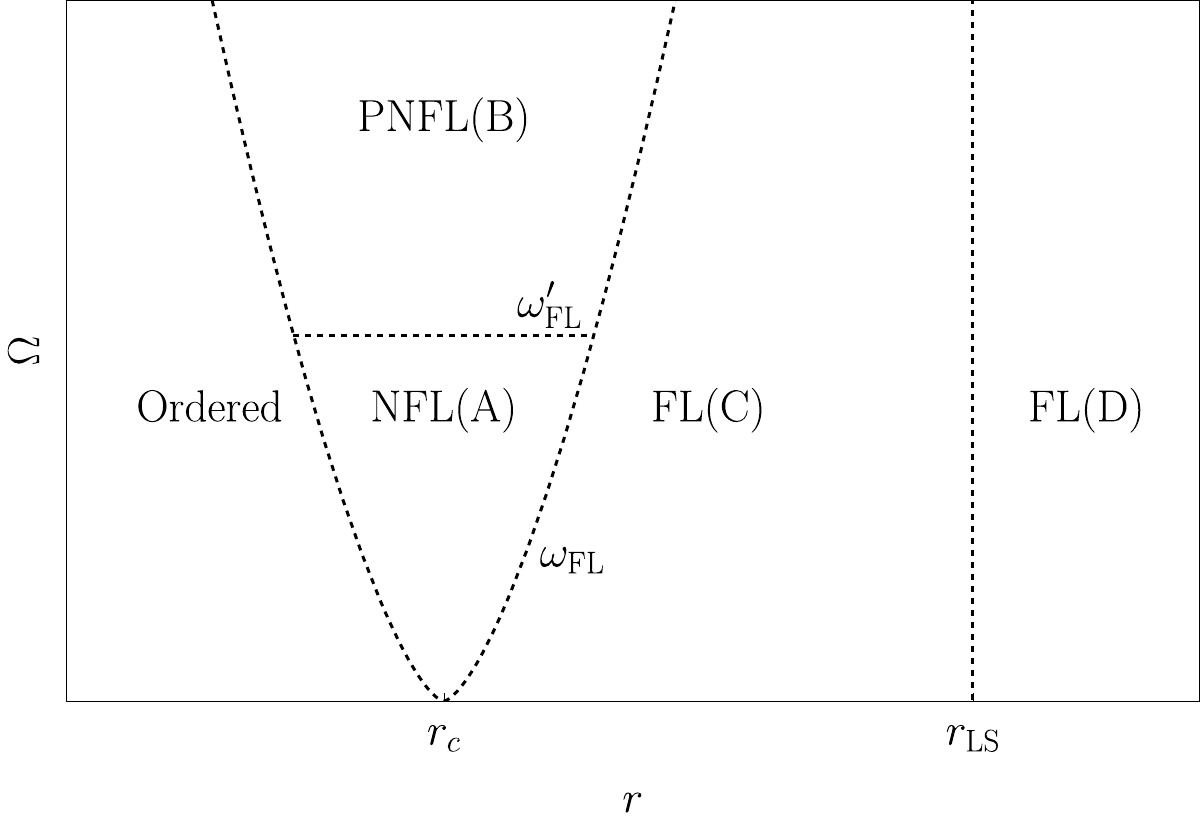}
  \caption{Phase diagram near the Ising-nematic quantum critical point in terms of external frequency $\Omega$ and the tuning parameter $r$. The quantum critical point corresponds to $m_b^2=r-r_c=0$. When $r<r_c$ the system orders at $T=0$. The regimes of interest are labelled A,B,C,D. A is the quantum critical NFL regime with $z=3$ boson. B is a perturbative NFL where the boson still has $z=3$ but the fermion self energy $\Sigma(i\omega)$ is small compared to $\omega$. C is a Fermi liquid regime where the boson $\phi$ acquires a mass term $0<m_b\ll k_F$. D is a Fermi liquid regime where the boson mass $m_b\gtrsim k_F$.}\label{fig:pd}
\end{figure}

    To be concrete, we will use the model of a fermion $\psi$ coupled to a real scalar boson $\phi$ with Yukawa coupling $g$, which describes the Ising-Nematic quantum critical point (QCP):
    \begin{equation}\label{eq:Lagrangian}
      \mathcal{L}=\psi^\dagger(\partial_\tau+\hat{\varepsilon}_k)\psi+\frac{1}{2}\phi\left(\hat{q}^2+r\right)\phi+g\psi^\dagger \psi \phi\,.
    \end{equation} Here $\hat{\varepsilon}_k$ and $\hat{q}^2$ are differential operators that correspond to the dispersion. We work in the unit where boson velocity $v_B=1$. The bare $\partial_\tau^2$ dynamics of the boson term is dropped because it is irrelevant compared to the generated Landau damping. Our calculation is not sensitive to the form factor of the Yukawa coupling, and we have set it to one for simplicity. The theory \eqref{eq:Lagrangian} will be handled using the Migdal-Eliashberg framework, which can be formally justified by the Yukawa-SYK model \cite{Iesterlis2021,HGuo2022a,EEAldape2022,ZDShi2022,ZDShi2023} or  the small-$(z_b-2)$-large-$N$ expansion \cite{DFMross2010}. We will use $r$ as a tuning parameter to access different regimes of the system as shown in Fig.~\ref{fig:pd}. The QCP is reached when $r=r_c$ and the boson is massless. Regime A is the NFL regime where the boson has dynamical exponent $z=3$ and the fermion self-energy is $\Sigma(i\omega)\propto \omega^{2/3}\gg \omega$. Regime B is the perturbative NFL (PNFL) regime where the scaling is the same as regime A but $\Sigma(i\omega)\ll \omega$. Regime C is the FL regime where the boson becomes massive with mass $m_b^2=r-r_c$. In regime C $m_b\ll k_F$ so the boson still mediates small-angle scattering. Regime D is also the FL regime but the boson mass $m_b$ becomes comparable to $k_F$ so the boson mediates large-angle scattering. Explicit calculations will be performed for regions A,B,C and qualitative results will be presented for regime D. We also note that in \cite{prlpaper} we showed that the Eliashberg theory is unstable in the NFL regime (A), but in this paper we will nevertheless calculate its transport properties and ignore the instability. We will perform calculations at $T=0$ and ignore the thermal fluctuations to extrapolate $\Omega$ scalings to finite $T$ to obtain scalings in $T$.

    The plan of the paper is as follows. In Sec.~\ref{sec:fluctuation} we will review the fluctuation spectrum of the 2+1D critical FS which is discussed in the companion paper \cite{prlpaper}. In particular, we will extend the analysis in \cite{prlpaper} to calculate the non-soft eigenvalues which is crucial for optical conductivity. In Sec.~\ref{sec:conductivity} we apply the formalism to calculate the optical conductivity. In Sec.~\ref{sec:hydro}, we apply the formalism to hydrodynamics and compute the viscosity.

\section{The Fluctuation Spectrum of the 2+1D critical FS}\label{sec:fluctuation}

In this section, we briefly review the formalism developed in the companion paper \cite{prlpaper}. In \cite{prlpaper} we mainly focused on the soft modes and here extend it to study the nonzero eigenvalues. Within the Migdal-Eliashberg theory, the Schwinger-Dyson equations for the theory \eqref{eq:Lagrangian} is given by a set of translational invariant solutions $G(x_1,x_2)=G(x_1-x_2)$, $D(x_1,x_2)=D(x_1-x_2)$ where $G$ and $D$ are the fermion and the boson Green's functions respectively, given by:
\begin{equation}\label{eq:MET}
      \begin{split}
         G(i\omega,\vec{k}) & =\frac{1}{i\omega-\xi_{\vec{k}}-\Sigma(i\omega,\vec{k})}\,, \\
         D(i\Omega,\vec{k})  &= \frac{1}{\vn{q}^2+m_b^2-\Pi(i\Omega,\vec{q})}\,, \\
         \Sigma(\tau,\vec{r})  & =g^2 G(\tau,\vec{r})D(\tau,\vec{r})\,, \\
         \Pi(\tau,\vec{r})  &=-g^2 G(\tau,\vec{r})G(-\tau,-\vec{r})\,.
\end{split}
\end{equation}
The object we study is the Bethe-Salpeter kernel $K_\text{BS}$, which can be defined through the bilinear fluctuation around the saddle point \eqref{eq:MET}:
\begin{equation}\label{eq:KBS}
      S[\delta G]=\frac{1}{2}\int_{x_1,x_2,x_3,x_4} \delta G(x_2,x_1)K_\text{BS}(x_1,x_2;x_3,x_4)\delta G(x_3,x_4)\,.
\end{equation} Here $\delta G(x_1,x_2)$ is the fluctuation of the fermion bilinear $-T_\tau\braket{\psi(x_1)\psi^\dagger(x_2)}$ around the SD equation solution \eqref{eq:MET}. Here $x_1, x_2$ etc. denote spacetime indices. The utility of $K_\text{BS}$ is to compute the linear response of fermion bilinears. For example, given two operators $\hat{A}=\int_{x_1,x_2}A(x_1,x_2)\psi^\dagger(x_1)\psi(x_2)$ and $\hat{B}=\int_{x_3,x_4} B(x_3,x_4)\psi^\dagger(x_3)\psi(x_4)$, we have
\begin{equation}\label{}
  \braket{\hat{A}\hat{B}}=\int_{x_1,x_2,x_3,x_4} A(x_2,x_1)K_\text{BS}^{-1}(x_1,x_2;x_3,x_4)B(x_3,x_4)\,.
\end{equation} As we will see, an efficient way to evaluate this functional inverse is to use its spectral decomposition.

From \eqref{eq:KBS}, we see that $K_\text{BS}$ is a functional that acts on two-point functions. It is also convenient to fourier transform two-point functions in real space to momentum space using the center-of-mass (CoM) and relative 3-momenta,
\begin{equation}\label{}
\begin{split}
  F(k;p)&=\int\rd^3 x_1 \rd^3 x_2 F(x_1,x_2)\\
  &\times\exp(-ip\cdot (x_1+x_2)/2-ik\cdot (x_1-x_2))\,.
\end{split}
\end{equation} Because of translational symmetry, $p$ is conserved by $K_\text{BS}$. In this section we will work in the homogeneous limit $p=(i\Omega,0)$ and the retarded branch $\Omega>0$. For notational clarity we will suppress $p$ in $F(k;p)$ unless otherwise mentioned.

As discussed in \cite{prlpaper}, the correct operator to diagonalize is the kinetic operator $L$:
\begin{equation}\label{}
  L=K_\text{BS}\circ M-\Omega I,
\end{equation} where $I$ is the identity operator and $M$ is a functional whose action in the momentum space is diagonal:
\begin{equation}\label{}
       M[F](i\omega,\vec{k})=(iG(i\omega+i\Omega/2,\vec{k})-iG(i\omega-i\Omega/2,\vec{k}))F(i\omega,\vec{k})\,.
\end{equation} The kinetic operator $L$ is symmetric under the following inner product:
  \begin{equation}\label{eq:innerprod}
     \begin{split}
       &\braket{A|B}=\int\frac{\rd\omega\rd^2\vec{k}}{(2\pi)^3}A(i\omega,\vec{k})\\
       &\times(iG(i\omega+i\Omega/2,\vec{k})-iG(i\omega-i\Omega/2,\vec{k}))
       B(i\omega,\vec{k})\,.
     \end{split}
     \end{equation}

From now on we assume the Fermi surface is circular, and we utilize the rotation symmetry to consider $L$ restricted to hangular harmonics number $m$ i.e. $F(k)=F(i\omega,\xi_k)e^{im\theta_k}$. The kinetic operator $L$ can be diagrammatically represented as the sum of density-of-states, Maki-Thompson and Aslamazov-Larkin diagrams. In \cite{prlpaper}, it is shown that $L_m$ can be handled using a double expansion in the dispersion $\xi$ and the boson momentum $\vn{q}$:
\begin{alignat}{4}\label{eq:Lexpand}
       L_m=&~L^{(0)}_m\quad+\quad&&~L^{(1)}_m\quad+\quad&&~L^{(2)}_m+\dots \nonumber\\[-2pt]
           &~~\verteq  &&~~\verteq  &&~~\verteq \nonumber\\[-4pt]
           &\delta_q^{0}L_m^{(0)}&&\delta_q^{0}L_m^{(1)}&&\delta_q^{0}L_m^{(2)}\\[-4pt]
           &~~+&&~~+&&~~+\nonumber\\[-4pt]
           &\delta_q^{1}L_m^{(0)}&&\delta_q^{1}L_m^{(1)}&&\delta_q^{1}L_m^{(2)}\nonumber\\[-4pt]
           &~~+&&~~+&&~~+\nonumber\\[-5pt]
           &~~~\vertdots&&~~~\vertdots&&~~~\vertdots\nonumber
     \end{alignat} Here the horizontal direction corresponds to the $\xi/(k_F v_F)$ expansion and the vertical direction corresponds to the $\vn{q}/k_F$ expansion.  In the large $k_F$ limit, the eigenvalues of $L_m$ are dominated by the leading order term $\delta_q^0 L_m^{(0)}$ which describes the forward scattering limit due to shape fluctuations of the FS. The exception is when zero modes occur and the effects of the higher order terms leads to be taken into account seriously, which has been done in \cite{prlpaper}. Here we are instead interested in the nonzero eigenvalues so it is sufficient to only consider $\delta_q^0 L_m^{(0)}$. The expression for $L_m$ in this limit is given by the sum of the following
\begin{widetext}
     \begin{equation}\label{eq:LMTT}
        \begin{split}
          \delta_q^0 L^{(0)}_{\text{MT+DOS},m}[F]&(i\omega,\xi)=g^2\int_{-\infty}^{\infty}\frac{\rd \omega'}{2\pi}\frac{\calN \rd \xi'}{2\pi}\int\frac{\rd\vn{q}}{k_F}D(\vn{q},i\omega-i\omega')\\
          &\times 2\left[iG(i\omega'+i\Omega/2,\xi')-iG(i\omega'-i\Omega/2,\xi')\right]\left[F(i\omega,\xi)-F(i\omega',\xi')\right]\,.
        \end{split}
        \end{equation}
        \begin{equation}\label{eq:LALT}
        \begin{split}
        \delta_q^0 L^{(0)}_{\text{AL},m}[F](i\omega_1,\xi_1)&=-g^4(2\pi)^2\left(1+(-1)^m\right)\calN^3\sgn\omega_1\int_{-\infty}^{\infty}\frac{\rd \nu}{2\pi}\frac{\rd \omega_2}{2\pi}\frac{\rd \xi_2}{2\pi}\int_0^\infty \frac{\rd \vn{q}}{k_F^2\vn{q}}\\
        &\times D(\vn{q},i\nu+i\Omega/2)D(\vn{q},i\nu-i\Omega/2) \sgn \omega_2 \theta(|\omega_1|-|\nu|)\theta(|\omega_2|-|\nu|)\\
        &\times i(G(i\omega_2+i\Omega/2,\xi_2)-G(i\omega_2-i\Omega/2,\xi_2))F(i\omega_2,\xi_2)\,.
        \end{split}
        \end{equation}
\end{widetext} Here $\calN=k_F/(2\pi v_F)$ is the density of states near the FS.

It will be convenient to write the function $F$ in a polynomial basis in $\xi$, as the following
\begin{equation}\label{eq:Fpoly}
  F(i\omega,\xi)=F_0(i\omega)+\frac{\xi}{A(i\Omega)}F_1(i\omega)+\left(\frac{\xi}{A(i\Omega)}\right)^2 F_2(i\omega)+\dots
\end{equation} Here $A(i\Omega)=i\Omega-\Sigma(i\Omega)$. The normalization $\xi$ by $A(i\Omega)$ makes the matrix element of $L_m$ of the equal scaling when acted on different powers of $\xi$. The hidden assumption of using the ansatz \eqref{eq:Fpoly} is that we are interested in smooth function of $\xi$, which is satisfied by the physical observables such as density or current. We will use the notation $\calH_n$ to denote the Hilbert space of $n$-th monomial in $\xi$.

The structure of $\delta_q^{(0)}L_m^{(0)}$ is fairly simple. In Eq.\eqref{eq:LMTT} the term proportional to $F(i\omega,\xi)$ is diagonal in the $(i\omega,\xi)$ domain, and in the rest of \eqref{eq:LMTT} and \eqref{eq:LALT} the result of the integral is independent of $\xi$. Therefore $\delta_q^{(0)}L_m^{(0)}$ has the following uppertriangular structure when written as blocks in $\calH_n$:
\begin{equation}\label{}
  \delta_q^{0}L_m^{(0)}=\begin{pmatrix}
              \left. \delta_q^{0}L_m^{(0)}\right|_{\calH_0} & \star & \star & \\
              0 & \left.\delta_q^{0}L_m^{(0)}\right|_{\calH_1} & 0 & \dots\\
              0 & 0 & \left.\delta_q^{0}L_m^{(0)}\right|_{\calH_2} & \\
              &  \vdots & & \ddots
            \end{pmatrix}\,.
\end{equation}

The diagonal blocks with $n\geq 1$ has the form
\begin{widetext}
\begin{equation}\label{eq:Lambdaomega}
  \begin{split}
          \left.\delta_q^0 L^{(0)}_{m}\right\vert_{\calH_{n\geq 1}}[F]&(i\omega,\xi)=g^2F(i\omega,\xi)\int_{-\infty}^{\infty}\frac{\rd \omega'}{2\pi}\frac{\calN \rd \xi'}{2\pi}\int\frac{\rd\vn{q}}{k_F}D(\vn{q},i\omega-i\omega')\times 2\left[iG(i\omega'+i\Omega/2,\xi')-iG(i\omega'-i\Omega/2,\xi')\right]\\
          &=\underbrace{\left[i\Sigma(i\omega+i\Omega/2)-i\Sigma(i\omega-i\Omega/2)\right]}_{\Lambda_{\omega}}F(i\omega,\xi)\,.
        \end{split}
\end{equation}
\end{widetext} Here in the second line we have used the Eliashberg equations \eqref{eq:MET} to evaluate the integral. This implies all eigenvalues of the $n\geq 1$ blocks are labelled by $\Lambda_\omega$ as defined in \eqref{eq:Lambdaomega}.

The remaining problem is to obtain the eigenvalues of the zeroth block $\left.\delta_q^0 L_m^{(0)}\right\vert_{\calH_0}$. In this block the $F=F(i\omega)$ is a function of frequency only.  From now on we will need to know the details of the system in different regimes.

\subsection{The NFL regime (A) and the perturbative NFL regime (B)}

    In the NFL (A) and PNFL (B) regimes, the solution of Eq.\eqref{eq:MET} is given by \cite{HGuo2022a}
\begin{equation}\label{}
  \Sigma(i\omega)=-ic_f|\omega|^{2/3}\sgn \omega\,,\quad c_f=\frac{g^2}{2\sqrt{3}\pi v_F \gamma^{1/3}}\,,
\end{equation}
\begin{equation}\label{}
  D(i\Omega,\vec{q})=\frac{1}{\vn{q}^2+\gamma|\Omega|/\vn{q}}\,,\quad \gamma=\frac{\calN g^2}{v_F}\,.
\end{equation} Using this to evaluate the $\vn{q}$ integrals in Eqs.\eqref{eq:LMTT} and \eqref{eq:LALT}, we obtain
\begin{widetext}
\begin{equation}\label{eq:LMT0}
          \delta_q^0L_{\text{MT+DOS},m}^{(0)}[F](i\omega)=\frac{2}{3}c_f\int_{-\Omega/2}^{\Omega/2}\rd \omega' \frac{1}{|\omega-\omega'|^{1/3}}\left[F(i\omega)-F(i\omega')\right]\,.
    \end{equation}
    \begin{equation}\label{eq:LAL0}
        \begin{split}
          &\delta_q^0L_{\text{AL},m}^{(0)}[F](i\omega_1)=-\frac{2}{3}c_f\sgn \omega_1\frac{1+(-1)^m}{2}\int_{-\Omega/2}^{\Omega/2}\rd \omega_2\int_{-\Omega/2}^{\Omega/2}\rd\nu\theta(|\omega_1|-|\nu|)\theta(|\omega_2|-|\nu|)\\
          &\times \frac{1}{|\nu^2-\Omega^2/4|^{1/3}\left(|\nu+\Omega/2|^{2/3}+|\nu-\Omega/2|^{2/3}+|\nu^2-\Omega^2/4|^{1/3}\right)}\sgn \omega_2 F(i\omega_2)\,.
        \end{split}
    \end{equation}
\end{widetext} Because of the $\xi$-integral the range of $\omega$ has been bounded to $[-\Omega/2,\Omega/2]$. We observe that Eqs.\eqref{eq:LMT0} and \eqref{eq:LAL0} has a particle-hole symmetry $F(i\omega)\to F(-i\omega)$, implying that the eigenvalues can be considered separately for odd and even sectors, i.e. $F(-i\omega)=P F(i\omega)$, $P=\pm 1$. It is easy to diagonalize \eqref{eq:LMT0} and \eqref{eq:LAL0} numerically, the result can be written as
\begin{equation}\label{}
  \lambda^m_\alpha=\frac{2}{3}c_f \Omega^{2/3}\alpha\,,
\end{equation} where $\alpha$ is a dimensionless factor. Our numerical result is summarized in Table.~\ref{tab:L0Spec}.
\begin{table}
      \centering
      \begin{tabular}{|c|c|c|c|}
      \hline
      $P$ & $(-1)^m$ & Discrete Spectrum & Continuum Spectrum \\
      \hline
      \multirow{2}{*}{\centering 1} & 1  & \multirow{2}{*}{$\alpha=0$} & \multirow{2}{*}{$\alpha\in\left[1.386,1.890\right]$} \\
      \cline{2-2}
       & -1 & & \\
       \hline
       \multirow{2}{2em}{\centering -1} & 1 & $\alpha=0.856,1.411,1.484$ & $\alpha\in\left[1.498,1.890\right]$ \\
       \cline{2-4}
       & {\centering -1} & $\alpha=1.226,1.449,1.491$ & $\alpha\in\left[1.499,1.890\right]$\\
       \hline
    \end{tabular}
      \caption{Spectrum of $\left.\delta_q^0L_m^{(0)}\right\vert_{\calH_0}$ in regimes A and B in different sectors defined by $P$ and $(-1)^m$. Numerical values are obtained through a 5000 by 5000 discretization of Eqs.\eqref{eq:LMT0} and \eqref{eq:LAL0}.}\label{tab:L0Spec}
    \end{table}
\begin{table}
      \centering
      \begin{tabular}{|c|c|c|}
      \hline
      $(-1)^{m}$  & Discrete Spectrum & Continuum Spectrum \\
      \hline
      {\centering 1}   & {$\alpha=0.856$} & {$\alpha\in\left[1.386,1.890\right]$} \\
       \hline
       {\centering -1}  & {$\alpha=1.226$} & {$\alpha\in\left[1.386,1.890\right]$} \\
       \hline
    \end{tabular}
      \caption{The approximate non-soft spectrum of $L_m$ in regimes A and B after breaking particle-hole symmetry.}\label{tab:L0Spec2}
    \end{table}

    For the $P=1$ sector, we found a zero mode which is studied in detail in the companion paper \cite{prlpaper}. Apart from that, we found a continuum spectrum with $\alpha\in [1.386,1.890]$. In the $P=-1$ sector, we also found 3 discrete modes together with a continuum as shown in Table.~\ref{tab:L0Spec}.

    However, the particle-hole symmetry only holds for the leading order term $L_m^{(0)}$. Generically $L_m^{(n)}$ has signature $(-1)^n$ under particle-hole transformation $(\omega,\xi)\to(-\omega,-\xi)$, so the symmetry is broken for $L_m$. Therefore, after putting back the higher perturbations $P=\pm 1 $ modes will hybridize and the spectra merge into one, with some discrete modes buried into the continuum. So the correct approximate spectrum should be given by Table.~\ref{tab:L0Spec2}.

\subsection{The FL regime with small-angle scattering (C)}

    The boson propagator is
\begin{equation}\label{}
  D(i\Omega,\vec{q})=\frac{1}{\vn{q}^2+m_b^2+\gamma|\Omega|/\vn{q}}\,.
\end{equation} The FL regime is characterized by the condition that the Landau damping only enters perturbatively, with the criterion $|\Omega|\ll \omega_\text{FL}=m_b^3/\gamma$. The fermion self-energy is then evaluated perturbatively in $1/m_b$, with the result
\begin{equation}\label{}
        \Sigma(i\omega)=(-i\omega)c_f'\left(\pi+\frac{|\omega|}{\omega_{\text{FL}}}\ln\left(\frac{|\omega|}{\omega_{\text{FL}}}\right)\right),
\end{equation} where $c_f'=g^2\calN/(2\pi k_F m_b)$.

    With this information, we can again perform the $q$-integral in \eqref{eq:LMTT} and \eqref{eq:LALT} to obtain
\begin{widetext}
\begin{equation}\label{eq:LMT02}
   \delta_q^0L_{\text{MT+DOS},m}^{(0)}[F](i\omega)=c_f' \int_{-\Omega/2}^{\Omega/2}\rd \omega' \left(\pi+\frac{2|\omega-\omega'|}{\omega_{\text{FL}}}\ln\left(\frac{|\omega-\omega'|}{\omega_{\text{FL}}}\sqrt{e}\right)\right)(F(i\omega)-F(i\omega'))\,,
\end{equation}
\begin{equation}\label{eq:LAL02}
\begin{split}
   &\delta_q^0L_{\text{AL},m}^{(0)}[F](i\omega)=2\frac{c_f'}{\omega_{\text{FL}}}\frac{1+(-1)^{m}}{2}\int_{-\Omega/2}^{\Omega/2}\rd \nu\int_{-\Omega/2}^{\Omega/2} \rd \omega_2 \sgn\omega_1 \theta(|\omega_1|-|\nu|) \theta(|\omega_2|-|\nu|) \\ &\frac{|\nu+\Omega/2|\ln\left(\frac{\gamma|\nu+\Omega/2|}{m_b^3}\sqrt{e}\right)-|\nu-\Omega/2|\ln\left(\frac{\gamma|\nu-\Omega/2|}{m_b^3}\sqrt{e}\right)}{|\nu+\Omega/2|-|\nu-\Omega/2|}\sgn\omega_2 F(\omega_2)\,.
\end{split}
\end{equation}
\end{widetext}

\begin{figure}
  \centering
  \includegraphics[width=0.95\columnwidth]{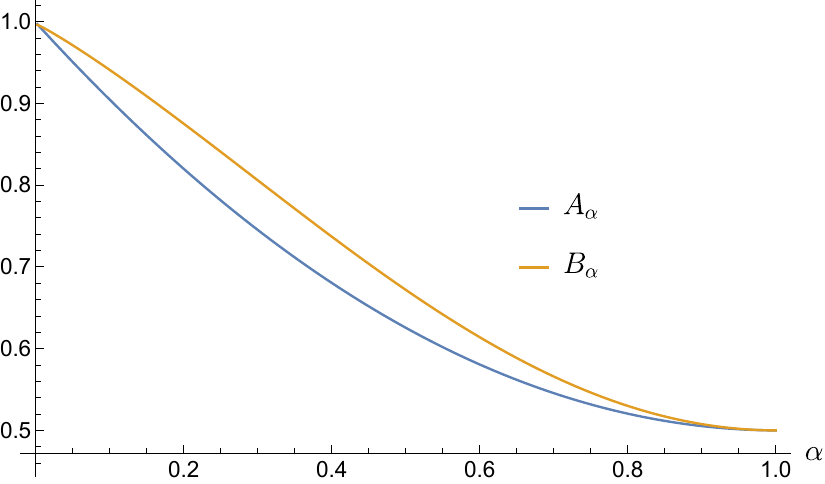}
  \caption{The functions $A_\alpha$ and $B_{\alpha}$ parameterizing the continuum spectrum in regime C.}\label{fig:AaBa}
\end{figure}

    Similar to the regimes A and B, we can numerically diagonalize Eqs. \eqref{eq:LMT02} and \eqref{eq:LAL02}, and we found the result can be well fitted by the function
\begin{equation}\label{eq:lambdaFL}
  \lambda_{\alpha}=c_f' A_\alpha \frac{\Omega^2}{\omega_{\text{FL}}}\ln\left(\frac{B_\alpha \Omega}{\omega_{\text{FL}}}\right)\,.
\end{equation} The parameters $A_\alpha$ and $B_\alpha$ are summarized in Table.~\ref{fig:L0sepcFL} and Fig.~\ref{fig:AaBa}. We found three discrete modes together with a continuum (the zero mode is already excluded). The parameters of the discrete modes are listed in Table.~\ref{fig:L0sepcFL} and the continuum parameters are plotted in Fig.~\ref{fig:AaBa} as a function of $\alpha\in[0,1]$.
\begin{widetext}
\onecolumngrid
\begin{table}
  \centering
  \begin{tabular}{|c|c|c|c|c|c|}
    \hline
      & $(-1)^m$ & \multicolumn{3}{|c|}{Discrete } & Continuum \\
      \hline
    \multirow{2}{*}{\centering $(A_\alpha,B_\alpha)$} & 1 & (0.995,0.981) & (2.000,1.407) & (0.998,0.994) & \multirow{2}{*}{Fig.~\ref{fig:AaBa}} \\
    \cline{2-5}
     & -1 & (0.995,0.981) & (1.219,1.076) & (0.999,0.999) &   \\
    \hline
  \end{tabular}
  \caption{The approximate spectrum of $L_m$ in regime C. The first discrete mode is approximately even and the other two modes are approximately odd under particle-hole symmetry.}\label{fig:L0sepcFL}
\end{table}
\twocolumngrid
\end{widetext}

\subsection{The FL regime with large-angle scattering (D)}

    In this regime, the form of the kinetic operator $L_m$ is complicated because the boson momentum $\vn{q}$ can be as large as $2k_F$. Therefore, we do not have analytical control as we did in regimes A,B,C. However, from the analysis above we see that the nonzero eigenvalues of $L_m$ is not suppressed compared to the self-energy scaling by the small-angle scattering. Therefore, we expect the nonzero eigenvalues of $L_m$ should scale similarly as the self-energy and be qualitatively similar to regime C.

\section{Optical Conductivity}\label{sec:conductivity}

In this section, we apply the eigenvalues we obtained to calculate the optical conductivity at the homogeneous limit $\vec{q}=0$. Rewriting the Kubo formula derived in \cite{HGuo2022a}, the optical conductivity before analytic continuation can be written as
\begin{equation}\label{}
  \sigma(i\Omega)=\frac{2\pi e^2}{\Omega}\braket{v_k|\frac{1}{L+\Omega}|v_k}\,.
\end{equation} We can expand it using eigenvalues of $L$, and we obtain
\begin{equation}\label{}
  \sigma(i\Omega)=\frac{2\pi e^2}{\Omega}\sum_{i} \frac{\braket{v_k|i}\braket{i|v_k}}{\Omega+\lambda_i}\,.
\end{equation} Therefore, we need to compute the overlap of velocity with the eigenfunctions using inner product \eqref{eq:innerprod}. The details depend on the FS shape and dispersion, which we discuss below.

\subsection{Circular FS with non-Parabolic Dispersion}

We first consider a circular FS with non-Parabolic dispersion. We expand $\vn{k}$ as a function of $\xi_k$ around the Fermi level:
\begin{equation}\label{eq:kexpand}
      \vn{k}=k_F+\frac{\xi_k}{v_F}-\frac{\kappa}{2}\frac{\xi_k^2}{k_Fv_F^2}+\frac{\zeta}{2}\frac{\xi_k^3}{k_F^2v_F^3}+\mathcal{O}(\xi_k^4)\,.
\end{equation} $\kappa$ and $\zeta$ are dimensionless numbers. In a Galilean invariant system, $\kappa=\zeta=1$. The velocity is then
\begin{equation}\label{}
  v_k=v_F+\frac{\kappa \xi_k}{k_F}+\frac{2\kappa^2-3\zeta}{2k_F^2v_F}\xi_k^2\,.
\end{equation} We will perform calculation to order $\mathcal{O}(\xi^2_k)$.

With the rotation symmetry, we only need to consider eigenvectors of $L_1$. $L_1$ contains a zero mode which is the momentum $\ket{k}$. We decompose the velocity $\ket{v_k}=\ket{v_k^\parallel}+\ket{v_k^\perp}$ which is the projection along the momentum and the orthogonal complement.
The overlap between the velocity $\ket{v_k}$ and the momentum is
    \begin{equation}\label{}
      \frac{\braket{v_k|k}\braket{k|v_k}}{\braket{k|k}}=v_F^2\braket{1|1}+\frac{2\kappa^2+2\kappa-3\zeta-1}{k_F^2}\braket{\xi|\xi}\,.
    \end{equation} We note that at low-energy when $\braket{\xi|\xi}$ can be neglected, the overlap is one hundred percent. This is due to the circular geometry of the FS.

    The  orthogonal complement is
    \begin{equation}\label{eq:vkperp}
      \ket{v_k^\perp}=\ket{v_k}-\ket{k}\frac{\braket{k|v_k}}{k|k}
      \approx (\kappa-1)\frac{\ket{\xi}}{k_F}\,.
    \end{equation} Because $\ket{v_k^\perp}\propto \ket{\xi}$, its eigenvalues are given simply by $\Lambda_\omega$ as defined in Eq.\eqref{eq:Lambdaomega}.

    The optical conductivity can be decomposed into two parts
    \begin{equation}\label{}
      \sigma(\omega)=\sigma_\text{D}(\omega)+\sigma_\text{i}(\omega)\,.
    \end{equation} The Drude contribution $\sigma_\text{D}$ is due to the projection of $\ket{v_k}$ along the momentum
    \begin{equation}\label{eq:sigmaD_circ}
      \sigma_\text{D}(i\Omega)=\frac{e^2\calN v_F^2}{2}\frac{1}{\Omega}\left[1+\left(2\kappa^2+2\kappa+3\zeta-1\right)\frac{\braket{\xi|\xi}/\braket{1|1}}{v_F^2 k_F^2}\right]\,,
    \end{equation}  where
    \begin{equation}\label{}
      \frac{\braket{\xi|\xi}}{\braket{1|1}}=\frac{1}{\Omega}\int_{-\Omega/2}^{\Omega/2}\rd \omega \frac{A(i\omega+i\Omega/2)^2+A(i\omega-i\Omega/2)^2}{2}\,.
    \end{equation}  We recall that $A(i\omega)=i\omega-\Sigma(i\omega)$.

    The incoherent conductivity arises from the current that is orthogonal to the momentum, i.e. \eqref{eq:vkperp}. We have
    \begin{equation}\label{eq:sigmai_circ}
    \begin{split}
      \sigma_i(i\Omega)&=\frac{e^2 \calN v_F^2}{2}\left(\frac{\kappa-1}{v_Fk_F}\right)^2\frac{1}{\Omega}\int_{-\Omega/2}^{\Omega/2}\rd \omega\int\frac{\rd \xi}{2\pi}\\
      &\times\frac{(iG(i\omega+i\Omega/2,\xi)-iG(i\omega-i\Omega/2,\xi))\xi^2}{\Omega+\Lambda_\omega}\\
      &=\frac{e^2 \calN v_F^2}{2}\left(\frac{\kappa-1}{v_Fk_F}\right)^2\frac{1}{\Omega}\int_{-\Omega/2}^{\Omega/2}\rd \omega\\&
      \times \frac{1}{2}\frac{A(i\omega+i\Omega/2)^2+A(i\omega-i\Omega/2)^2}{\Omega+\Lambda_\omega}\,,
    \end{split}
    \end{equation}

    We notice that in the Galilean invariant limit $\kappa=\zeta=1$, all the correction terms vanish and we are left with the Drude result $\sigma(i\Omega)=e^2\calN^2 v_F^2/(2\Omega)$.

    Next we can evaluate Eqs.\eqref{eq:sigmaD_circ} and \eqref{eq:sigmai_circ} in different regimes. We notice that in the NFL regime (A), $A(i\omega)\approx -\Sigma(i\omega)$ is dominated by the self-energy and $\Lambda_\omega\gg \Omega$. In other regimes $A(i\omega)\approx i\omega$ and $\Lambda_\omega\ll \Omega$. In the NFL (A) and the PNFL (B) regimes we neglect the effect of boson thermal mass and the assumed $\omega/T$ scaling.

    \begin{table}[tb!]
      \centering
      \begin{tabular}{|c|c|c|c|}
        \hline
         Regimes & NFL(A) & PNFL(B) & FL(C,D) \\
        \hline
        \multirow{3}{*}{{$\displaystyle\frac{\Re \sigma_D(\omega\gg T)}{e^2\calN^2 v_F^2}$}} & \multirow{3}{*}{$\displaystyle\frac{c_f^2|\omega|^{1/3}}{k_F^2 v_F^2}$} & \multirow{3}{*}{0} & \multirow{3}{*}{0}\\
        & & &\\
        &&&\\
        \hline
        \multirow{3}{*}{$\displaystyle\frac{\Re \sigma_D(\omega\ll T)}{e^2\calN^2 v_F^2}$} &
        \multirow{3}{*}{$\displaystyle\frac{c_f^2 T^{4/3}}{k_F^2v_F^2}\delta(\omega)$}
         & \multirow{3}{*}{$\displaystyle\frac{T^2}{k_F^2 v_F^2}\delta(\omega)$}  & \multirow{3}{*}{$\displaystyle\frac{T^2}{k_F^2 v_F^2}\delta(\omega)$}  \\
        &&&\\
        &&&\\
        \hline
        \multirow{3}{*}{{$\displaystyle\frac{\Re \sigma_i(\omega\gg T)}{e^2\calN^2 v_F^2}$}} &
        \multirow{3}{*}{$\displaystyle\frac{c_f|\omega|^{2/3}}{k_F^2 v_F^2}$} &
        \multirow{3}{*}{$\displaystyle\frac{c_f|\omega|^{2/3}}{k_F^2 v_F^2}$} &
        \multirow{3}{*}{$\displaystyle\frac{c_f'|\omega|^2\ln(|\omega|/\omega_\text{FL})}{\omega_\text{FL}k_F^2v_F^2}$}\\
        &&&\\
        &&&\\
        \hline
        \multirow{3}{*}{{$\displaystyle\frac{\Re \sigma_i(\omega\ll T)}{e^2\calN^2 v_F^2}$}} &
        \multirow{3}{*}{$\displaystyle\frac{c_f T^{2/3}}{k_F^2 v_F^2}$}&
        \multirow{3}{*}{$\displaystyle\frac{c_fT^{8/3}}{|\omega|^2k_F^2 v_F^2}$} &
         \multirow{3}{*}{$\displaystyle\frac{c_f'T^4\ln(T/\omega_\text{FL})}{\omega_\text{FL}|\omega|^2k_F^2v_F^2}$}\\
        &&&\\
        &&&\\
        \hline
      \end{tabular}
      \caption{Corrections to the real part of the optical conductivity of a circular FS in different regimes. The zeroth order Drude peak $\sigma_D^{(0)}(\omega)=\pi e^2\calN v_F^2 \delta(\omega)/2$ is subtracted. The numerical prefactors are neglected.}\label{tab:sigma_circ}
    \end{table}

     Our results are summarized in Table.~\ref{tab:sigma_circ} where the results have been continued to real time $i\Omega\to \omega+i0$.

     For the Drude peak term $\sigma_D$, we find a correction to the Drude weight at low frequency. At high frequencies, the correction to $\Sigma_D$ is purely imaginary in the PNFL and the FL regime, but there is a real component in the NFL regime.

     The incoherent conductivity $\sigma_i$ is nonzero in all four regimes. In the NFL regime A, the incoherent part shows $\omega/T$ scaling, but it is sub-leading to the Drude peak term. In the PNFL regime B and FL regimes C,D, because $\Sigma<\omega$ the resulting $\sigma_i$ does not show $\omega/T$ scaling.

     \subsection{Non circular FS}

     When the FS is not circular, the velocity $\ket{v_k}$ does not fully overlap with momentum $\ket{k}$ even at zero energy. We decompose it as
     \begin{equation}\label{}
       \ket{v_k}=\ket{v_k^\parallel}+\ket{v_{k,\xi=0}^\perp}+\ket{v_{k,\xi}^\perp}\,.
     \end{equation} Here $\ket{v_k^\parallel}$ is still the projection along the momentum, but the orthogonal complement has been further decomposed into a $\xi$-independent part $\ket{v_{k,\xi=0}^\perp}$, and a $\xi$-dependent part $\ket{v_{k,\xi}^\perp}$.

     Following this decomposition, $\ket{v_k^\parallel}$ and $\ket{v_{k,\xi}^\perp}$ still contributes to $\sigma_D$ and $\sigma_i$ respectively similar to the circular FS case, but the numerical prefactors are different because of different FS geometry.

     The interesting new contribution is from $\ket{v_{k,\xi=0}^\perp}$, which overlaps with the old-parity soft modes studied in the companion paper \cite{prlpaper} when the FS is not circular. It leads to a term that looks like a modified Drude peak
     \begin{equation}\label{eq:sigmaDp}
       \sigma_D'(i\Omega)=e^2\calN v_F^2 \frac{1}{\Omega+\lambda^\text{soft}_\text{odd}}\,.
     \end{equation}

     We summarize the result for the soft eigenvalues $\lambda^\text{soft}_\text{odd}$  in Table.~\ref{tab:soft}. The results for regimes A,B,C are calculated in the companion paper \cite{prlpaper}. Regime D, which involves large-angle scattering, is not as analytically tractable as regimes A,B,C. However, since D is deep in the FL, we quote the result of Ref.~\cite{PJLedwith2019} which calculated the soft mode eigenvalues using classical Boltzmann equation, and it agrees with qualitative arguments presented in \cite{prlpaper}.

   Next, we apply the soft mode results in Table.~\ref{tab:soft} to Eq.\eqref{eq:sigmaDp}, and the resulting $\sigma_D'(\omega)$ is summarized in Table.~\ref{tab:sigmaDp}. We have assumed the absence of boson thermal mass and so the soft mode eigenvalues are assumed to satisfy $\omega/T$ scaling.

 \begin{widetext}
  \onecolumngrid
  \centering
  \begin{table}
  \centering
  \begin{tabular}{|c|c|c|c|c|c|}
    \hline
    Regime &  $\lambda^{\text{soft}}_{\text{even }m}$ & $\lambda^\text{soft}_{\text{odd }m}$ (Convex FS) & $\lambda^\text{soft}_{\text{odd }m}$ (Concave FS) \\ \hline
    A & $m^2(c_f\Omega^{2/3})^2/(k_Fv_F)$ & $m^2(m^2-1)^2(c_f \Omega^{2/3})^{4}/(k_F v_F)^3$ & $m^2(c_f\Omega^{2/3})^2/(k_Fv_F)$\\ \hline
    B & $m^2(c_f\Omega^{2/3})^2/(k_Fv_F)$  & $m^2(m^2-1)^2(c_f \Omega^{2/3})^{2}\Omega^2/(k_F v_F)^3$ & $m^2(c_f\Omega^{2/3})^2/(k_Fv_F)$ \\\hline
    C & $m^2m_b^2c_f' \Omega^2/(k_F^2\omega_\text{FL})$  & $m^2(m^2-1)^2m_b^2c_f'\Omega^4/(k_F^4v_F^2\omega_\text{FL})$ & $m^2m_b^2c_f' \Omega^2/(k_F^2\omega_\text{FL})$ \\\hline
    D & $c_f' \Omega^2\ln m/\omega_\text{FL}$  &  $c_f' \Omega^4m^4\ln m /(k_F^2v_F^2\omega_\text{FL})$ & $c_f' \Omega^2\ln m/\omega_\text{FL}$ \\
    \hline
  \end{tabular}
  \caption{The dissipative part of the soft mode eigenvalues in different regimes at $T=0$. $m$ is the angular harmonic number of a circular FS, and should be interpreted as the eigenvalue of the angular Laplacian for a general FS. Regimes A,B,C are computed in the companion paper \cite{prlpaper}. Regime D is from \cite{PJLedwith2019}.}\label{tab:soft}
\end{table}

    \begin{table}
      \centering
      \begin{tabular}{|c|c|c|c|c|c|}
        \hline
          & FS geometry & A & B & C & D \\
          \hline
        $\displaystyle\frac{\Re\sigma_D'(\omega\gg T)}{e^2\calN v_F^2}$ & \multirow{2}{*}{Convex} & $\displaystyle\frac{c_f^4|\omega|^{2/3}}{(k_Fv_F)^3}$  &  $\displaystyle\frac{c_f^2|\omega|^{4/3}}{(k_F v_F)^3}$ & $\displaystyle\frac{m_b^2}{k_F^2}\frac{c_f' |\omega|^2}{k_F^2 v_F^2 \omega_\text{FL}}$  & $\displaystyle\frac{c_f' |\omega|^2}{k_F^2 v_F^2 \omega_\text{FL}}$  \\
        \cline{1-1} \cline{3-6}
        $\displaystyle\frac{\Re\sigma_D'(\omega\ll T)}{e^2\calN v_F^2}$ & &  $\displaystyle\frac{c_f^4T^{8/3}}{(k_Fv_F)^3|\omega|^2}$ &
        $\displaystyle\frac{c_f^2T^{10/3}}{(k_F v_F)^3|\omega|^2}$  &
        $\displaystyle\frac{m_b^2}{k_F^2}\frac{c_f' T^4}{k_F^2 v_F^2 \omega_\text{FL}|\omega|^2}$  &
        $\displaystyle\frac{c_f' T^4}{k_F^2 v_F^2 \omega_\text{FL}|\omega|^2}$  \\
        \hline
        $\displaystyle\frac{\Re\sigma_D'(\omega\gg T)}{e^2\calN v_F^2}$ & \multirow{2}{*}{Concave} &
        $\displaystyle\frac{c_f^2}{k_F v_F|\omega|^{2/3}}$ & $\displaystyle\frac{c_f^2}{k_F v_F|\omega|^{2/3}}$  &
         $\displaystyle \frac{m_b^2}{k_F^2}\frac{c_f'}{\omega_\text{FL}}$&  $\displaystyle\frac{c_f'}{\omega_\text{FL}}$ \\
        \cline{1-1} \cline{3-6}
        $\displaystyle\frac{\Re\sigma_D'(\omega\ll T)}{e^2\calN v_F^2}$ & & $\displaystyle\frac{c_f^2 T^{4/3}}{k_F v_F|\omega|^{2}}$  &  $\displaystyle\frac{c_f^2 T^{4/3}}{k_F v_F|\omega|^{2}}$ & $\displaystyle\frac{m_b^2}{k_F^2}\frac{c_f' T^2}{|\omega|^2\omega_\text{FL}}$  &   $\displaystyle\frac{c_f' T^2}{|\omega|^2\omega_\text{FL}}$    \\
        \hline
      \end{tabular}
      \caption{The new Drude-like contribution for non-circular FS in different regimes. We assumed $\omega/T$ scaling for the soft mode eigenvalues.}\label{tab:sigmaDp}
    \end{table}
\twocolumngrid
\end{widetext}

    To summarize, the total optical conductivity is the sum of all three contributions
    \begin{equation}\label{}
      \sigma(\omega)=\sigma_D(\omega)+\sigma_D'(\omega)+\sigma_i(\omega).
    \end{equation}

  Our results agree with the recent calculation \cite{SLi2023} which calculated $\sigma(\omega)$ using perturbation theory in the FL regime and later extrapolated to the PNFL regime. The results in \cite{SLi2023} for the circular and the convex FS map to our incoherent conductivity $\sigma_i$ in regimes B and C. The results in \cite{SLi2023} for the concave FS, map to the our $\sigma_D'$ term in regimes B and C. However, because the calculation in \cite{SLi2023} assumes $\Sigma\ll \omega$, it was unable to access our regime A which is the true NFL regime.

\section{Hydrodynamics}\label{sec:hydro}

In this section, we discuss the implication of our formalism in hydrodynamics. In the hydrodynamic regime, the non-local quantities quickly relax through local collision and only zero modes or soft modes are left. These conserved and quasi-conserved quantities then can start propagating in space. As a result of this propagation, the current-field relation $\vec{j}=\sigma\vec{E}$ is no-longer given by a conductivity $\sigma$, but instead is described through the non-local conductivity $\sigma(\vec{p})$:
\begin{equation}\label{eq:sigmaq}
  \sigma(\vec{p})=\frac{ne^2}{2\pi\calN \vn{p}^2 \nu(\vec{p})}\,.
\end{equation} Here $n$ is the total fermion density, and $\nu(\vec{p})$ is the kinematic viscosity, which can be written as
\begin{equation}\label{eq:nuq}
  \nu(\vec{q})=\frac{n}{4\pi\calN^2 \Gamma_2(\vec{p})}\,,
\end{equation} and $\Gamma_2(\vec{p})$ is the effective scattering rate. The wavevector $\vec{p}$ should be understood as a typical wavevector of the external drive, such as inverse sample size.

\subsection{Derivation of a Boltzmann equation for the soft modes}

    To capture the non-local conductivity $\sigma(\vec{p})$, we need to allow a finite CoM momentum in our formalism, i.e. $p=(i\Omega,\vec{p})$. In this subsection, we discuss how a Boltzmann equation naturally emerges from the soft modes. Unless otherwise mentioned, we will work with the circular FS.

    Since the soft modes can now propagate in space, the eigenvalues of the kinetic operator $L$ or the Bethe-Salpeter kernel $K_\text{BS}$ is no longer an intrinsic property of the critical FS because it depends on the real space geometry of system. However, it is still useful to consider the operator
    \begin{equation}\label{}
      \mathcal{L}=K_\text{BS}\circ M\,.
    \end{equation} We expand $K_\text{BS}=W_\Sigma^{-1}-W_\text{MT}-W_\text{AL}$ into three sets of Feynmann diagram contributions \cite{HGuo2022a,prlpaper}. We obtain
    \begin{widetext}
    \begin{equation}\label{eq:Wsigmap}
      W_\Sigma^{-1}\circ M=G^{-1}(k+p/2)G^{-1}(k-p/2)\left[iG(k+\Omega/2)-iG(k-\Omega/2)\right]\,.
    \end{equation}
    \begin{equation}\label{eq:tLMT}
      \tilde{L}_{\text{MT}}[F](k)=-W_\text{MT}\circ M [F](k)=g^2\int\frac{\rd^3k'}{(2\pi)^3}D(k-k')\left[iG(k'+i\Omega/2)-iG(k'-i\Omega/2)\right]\left[-F(k')\right]\,.
    \end{equation}
    \begin{equation}\label{eq:tLAL}
    \begin{split}
      \tilde{L}_\text{AL}[F](k_1)&=-W_\text{AL}\circ M[F](k_1)=g^4 \int \frac{\rd^3 q\rd^3 k_2}{(2\pi)^6}G(k_1-q)\left(G(k_2-q)+G(k_2+q)\right)D(q+p/2)D(q-p/2)\\
      &\times \left[iG(k_2+i\Omega/2)-iG(k_2-i\Omega/2)\right]F(k_2)\,.
    \end{split}
    \end{equation}
    \end{widetext}

    The expression for the Maki-Thompson part \eqref{eq:tLMT} is identical to the homogeneous limit ($\vec{p}=0$). The Aslamazov-Larkin part \eqref{eq:tLAL} is slightly different because the boson propagator pair $D(q+p/2)D(q-p/2)$ knows about the finite CoM momentum. This $\vec{p}$ dependence reflects the fact that the boson can also propagate in space. However, within the Eliashberg approximation the bosons move much slower than the fermions, so we can ignore this effect and assume $\tilde{L}_\text{AL}$ also coincides with the homogeneous limit. The last part to take care of is Eq.\eqref{eq:Wsigmap}. We expand it to linear order in $\vec{p}$, and we obtain
    \begin{widetext}
    \begin{equation}\label{eq:WSigmap2}
      W_\Sigma^{-1}\circ M=\Omega+i\vec{v}_k\cdot\vec{p}+i\Sigma(i\omega+i\Omega/2)-i\Sigma(i\omega-i\Omega/2)+\delta\mathcal{L}\,,
    \end{equation} where $\vec{v}_k=\nabla_k \xi_k$ and
    \begin{equation}\label{}
      \delta\mathcal{L}=-\frac{i}{2}\vec{v}_k\cdot \vec{p}\left(\frac{G(i\omega+i\Omega/2,\vec{k})}{G(i\omega-i\Omega/2,\vec{k})}+\frac{G(i\omega-i\Omega/2,\vec{k})}{G(i\omega+i\Omega/2,\vec{k})}\right)\,.
    \end{equation}
    \end{widetext}
    Combining everything together, we obtain
    \begin{equation}\label{eq:calL2}
      \mathcal{L}=\Omega+i\vec{v}_k\cdot \vec{p}+L_\text{MT+DOS}+L_\text{AL}+\delta\mathcal{L}\,.
    \end{equation} Here the self-energy terms in \eqref{eq:WSigmap2} are combined with \eqref{eq:tLMT} to obtain $L_\text{MT+DOS}$, and we approximated $\tilde{L}_\text{AL}$ by $L_\text{AL}$.  To access the hydrodynamic regimes, we project the operator \eqref{eq:calL2} onto the subspace of soft modes. In \cite{prlpaper} we have shown that to leading order the eigenfunctions of the soft modes are functions of $\theta_k$, $\Omega$ and $\vec{p}$ only (since the projection is local, we suppress the $\Omega$ and $\vec{p}$ dependence):
    \begin{equation}\label{}
      \ket{F_\text{soft}}=F(\theta_k)\,,
    \end{equation} so the projected matrix element will be $\braket{H_\text{soft}|\mathcal{L}|F_\text{soft}}$ where $F$ and $H$ are two soft mode wavefunctions.

    Under this projection, $\delta\mathcal{L}$ becomes
    \begin{widetext}
    \begin{equation}\label{}
    \begin{split}
      \braket{H_\text{soft}|\delta\mathcal{L}|F_\text{soft}}&=\calN\int\frac{\rd \omega}{2\pi}\frac{\rd \xi}{2\pi}\int\rd \theta_k \left(-\frac{i}{2}v_k\vn{p}\cos_{\theta_{kp}}\right)\left[\frac{G(i\omega+i\Omega/2,\xi)}{G(i\omega-i\Omega/2,\xi)}+\frac{G(i\omega-i\Omega/2,\xi)}{G(i\omega+i\Omega/2,\xi)}\right]\\
      &\times[iG(i\omega+i\Omega/2,\xi)-iG(i\omega-i\Omega/2,\xi)]F(\theta_k)H(\theta_k)\,.
    \end{split}
    \end{equation}
    \end{widetext} To leading order, we can approximate $v_k=v_F$, and we perform the $\xi$-integral by picking up the residues. However, the $\xi$-residue of the product of the two brackets is zero, so to leading order $\delta\mathcal{L}=0$ after projection, so the projected operator $\mathcal{\bar{L}}$ is
    \begin{equation}\label{eq:calL3}
      \mathcal{\bar{L}}=\Omega+i\vec{v}_k\cdot \vec{p}+\underbrace{L_\text{MT+DOS}+L_\text{AL}}_{L}\,.
    \end{equation}  After interpreting $L$ as the collision operator, this is exactly the kernel of the Boltzmann equation.

    Therefore, the computation of $\sigma(\vec{p})$ reduces to solving the Boltzmann equation $\mathcal{\bar{L}}F=\text{Drive}$.  We expand $F$ into angular harmonics of $\theta_k$:
    \begin{equation}\label{}
      F=\sum_{m}\frac{e^{im(\theta_k-\theta_p)}}{2\pi} F_m(\Omega,\vec{p})\,.
    \end{equation} Since $e^{im\theta_k}$ diagonalizes $L$, we have
    \begin{widetext}
    \begin{equation}\label{}
      \mathcal{\bar{L}}[F]=\sum_{m}\frac{e^{im(\theta_k-\theta_p)}}{2\pi}\left[(\Omega+\lambda_m^\text{soft})F_m(\Omega,\vec{p})+\frac{i v_F\vn{p}}{2}\left(F_{m-1}(\Omega,\vec{p})+F_{m+1}(\Omega,\vec{p})\right)\right]\,.
    \end{equation}
    \end{widetext} Following the same analysis as the FL hydrodynamics \cite{PLedwith2019,QHong2020,SKryhin2023}, the non-local conductivity $\sigma(\vec{p})$ and the kinematic viscosity $\nu(\vec{p})$ are exactly given by Eqs.\eqref{eq:sigmaq} and \eqref{eq:nuq} respectively, and the effective scattering rate is given by a continuous fraction formula
    \begin{equation}\label{eq:contfrac}
      \Gamma_2(\vec{p})=\gamma_2+\frac{z^2}{\gamma_3+\frac{z^2}{\gamma_4+\dots}}\,,
    \end{equation} where $z=v_F\vn{p}/2$ and $\gamma_m=\Omega+\lambda_m^\text{soft}$. For the soft eigenvalues to be significant, we should work in the limit $\Omega\ll T$. Again we ignore the effects of boson thermal mass and assume the $\Omega$-scalings of $\lambda_m^\text{soft}$ carries over to scaling in $T$.

    For short lengthscales where $z$ becomes larger than the self-energy, $\Gamma_2$ is described by the ballistic limit $\Gamma_2=z$. At longer lengthscales, $\Gamma_2$ can enter the hydrodynamic or the tomographic regimes, which we analyze below:

    \subsection{Conventional Hydrodynamics Regime }

    The conventional hydrodynamic regime is the ultimate long-wavelength regime where $\vec{p}\to 0$. In this regime, \begin{equation}\label{}
      \Gamma_2=\Gamma_\text{hydro}=\gamma_2\,,
    \end{equation}
    and its scalings are summarized in Table.~\ref{tab:Gammaeff}.

    \subsection{Tomographic Regime}

    When we go to a shorter length scale $\vn{p}>p_*$ which will be specified later, we enter into the tomographic regime where all soft modes can propagate and produces an effective scattering rate whose scaling differs from that of each individual eigenvalue.

    We follow the analysis in \cite{QHong2020,SKryhin2023} to solve the recurrence equation for the continuous fraction
    \begin{equation}\label{eq:Gammam}
      \Gamma_m(\vec{p})=\gamma_m+\frac{z^2}{\gamma_{m+1}+\frac{z^2}{\gamma_{m+2+\dots}}}\,.
    \end{equation} We assume $\gamma_m$ oscillates between two functions $\gamma_o(m)$ and $\gamma_e(m)$ when $m$ is odd and even respectively. In \cite{QHong2020,SKryhin2023}, it is shown that we can rewrite Eq.\eqref{eq:Gammam} into a linear recurrence equation in terms of some auxiliary variables $u_m$, and then we can approximate the recurrence equation by a differential equation for $u_m=u(m)$:
    \begin{equation}\label{eq:upp}
      u''-\frac{\gamma_o'}{\gamma_o}u'-\frac{\gamma_o \gamma_e}{4z^2}u=0\,,
    \end{equation}   where prime means derivative with respect to $m$. Then $\Gamma_m$ is obtained from $u_m$ by
    \begin{equation}\label{}
      \Gamma_m=-\frac{2z^2}{\gamma_o}\frac{\rd\ln u}{\rd m}\,.
    \end{equation} The boundary condition for Eq.\eqref{eq:upp} is $\Gamma_m>0$ when $m\to\infty$.

    In regimes A,B,C and for circular or convex FS, we have $\gamma_e=\gamma m^2$ and $\gamma_o\approx\gamma' m^6$, and Eq.\eqref{eq:upp} becomes
    \begin{equation}\label{}
      u''-\frac{6}{m}u'-\frac{\gamma\gamma'}{4z^2}m^8 u=0\,.
    \end{equation} The solution can be expressed in terms of Bessel functions
    \begin{equation}\label{}
      u_m=g^{7/10}\left(I_{7/10}(g)-I_{-7/10}(g)\right),
    \end{equation} where $g=\frac{m^5\sqrt{\gamma\gamma'}}{10z}$ and the linear combination is selected to satisfy the boundary condition.
    The asymptotic behavior of $\Gamma_2$ can then be evaluated by expanding in large $z$, and we obtain
    \begin{equation}\label{eq:Gamma2Circ}
      \Gamma_2=\frac{\Gamma\left(\frac{3}{10}\right)}{2^{4/5}5^{2/5}\Gamma\left(\frac{7}{10}\right)}\frac{\gamma z^{3/5}}{(\gamma\gamma')^{3/10}}-\frac{4\gamma}{3}+\dots.
    \end{equation} Therefore, the length scale below which the tomographic transport becomes important is given by
    \begin{equation}\label{}
      z=\frac{\vn{p}v_F}{2}\gg \sqrt{\gamma\gamma'}\,,
    \end{equation} where $\gamma, \gamma'$ can be read off from Table.~\ref{tab:soft}. We can now apply Eq.\eqref{eq:Gamma2Circ} to regimes A,B,C, and the result is summarized in Table.~\ref{tab:Gammaeff} as $\Gamma_\text{tomographic}^\text{circular}$.
    The result of regime D has been obtained in Refs. \cite{QHong2020,SKryhin2023}. The momentum scaling $\Gamma_2(\vn{p})\propto \vn{p}^{3/5}$ is the consequence of small-angle scattering, which sets a clear difference from the $\vn{p}^{1/3}$ scaling due to large-angle scattering \cite{QHong2020,SKryhin2023}.

    We also attempt to extend to the case concave FS, by substituting $\gamma_o=\gamma m^2$ and $\gamma_e=\gamma' m^2$. Because rotation symmetry is broken, the calculation above does not actually apply, but we hope it might provide the correct scaling. Repeating the analysis above for regimes A,B,C, we find the solution of the differential equation is given by
    \begin{equation}\label{}
      u_m=\exp(-\frac{\sqrt{\gamma\gamma'}}{6z}m^3)\,,
    \end{equation} and
    \begin{equation}\label{}
      \Gamma_2=\sqrt{\frac{\gamma'}{\gamma}}z\,.
    \end{equation} This is similar to the ballistic limit where $\Gamma_2=z$, but the prefactor is renormalized. As for regime D, because the even-$m$ eigenvalue scales the same way as self-energy, we expect the system to directly crossover to the ballistic regime without the tomographic regime.
\begin{widetext}
\onecolumngrid
\begin{table}
  \centering
  \begin{tabular}{|c|c|c|c|}
    \hline
    Regimes & $\displaystyle\Gamma_\text{hydro}$ & $\displaystyle\Gamma_\text{tomographic}^\text{Circular}$ & $\displaystyle\Gamma_\text{tomographic}^\text{Concave}$\\
    \hline
    A & $\displaystyle \frac{c_f^2 T^{4/3}}{k_F v_F}$ & $\displaystyle (\vn{p}v_F)^{3/5}(k_F v_F)^{1/5}(c_f T^{2/3})^{1/5}$ &  $\displaystyle \calO(1)\vn{p}v_F$  \\
    \hline
    B & $\displaystyle \frac{c_f^2 T^{4/3}}{k_F v_F}$ & $\displaystyle (\vn{p}v_F)^{3/5}(k_F v_F)^{1/5}c_f^{4/5}T^{-1/15}$ & $\displaystyle \calO(1)\vn{p}v_F$ \\
    \hline
    C & $\displaystyle \frac{m_b^2}{k_F^2}\frac{c_f' T^2}{\omega_\text{FL}}$ & $\displaystyle (\vn{p} v_F)^{3/5}(k_F v_F)^{3/5} c_f'^{2/5} T^{1/5}\omega_\text{FL}^{-2/5}(m_b/k_F)^{4/5}$ & $\displaystyle \calO(1)\vn{p}v_F$ \\
    \hline
    D & $\displaystyle\frac{c_f' T^2}{\omega_\text{FL}}$ & $\displaystyle c_f'^{2/3}(v_F\vn{p})^{1/3}(k_F v_F)^{1/3}\omega_\text{FL}^{-2/3}T$ & N.A. \\
    \hline
  \end{tabular}
  \caption{Scalings of the effective scattering rate $\Gamma_2$ defined in \eqref{eq:nuq} in the hydrodynamic regime and the tomographic regime. In the hydrodynamic regime, the effective scattering rate equals $\lambda_2^\text{soft}$. For the circular Fermi surface, in the tomographic regime $\Gamma_2$ depends on the external wavevector $\vn{p}$ and scales nontrivially with temperature due to the propagation of the soft modes. For the concave FS, the tomographic regime looks similar to the ballistic regime where $\Gamma_2\propto \vn{p}v_F$, but the numerical prefactor can be renormalized.  }\label{tab:Gammaeff}
\end{table}
\twocolumngrid
\end{widetext}

\section{Conclusion}

    In this work, we applied the kinetic operator formalism to study the transport properties of various regimes of the critical Fermi surface, using the Ising-Nematic QCP as an example (see the phase diagram Fig.~\ref{fig:pd}). We obtained the operator spectrum of the critical FS in Sec.~\ref{sec:fluctuation} and applied it to optical conductivity in Sec.~\ref{sec:conductivity}. We obtained results for different regimes in the phase diagram (Fig.~\ref{fig:pd}) and different FS geometries as summarized in Tables.~\ref{tab:sigma_circ} and \ref{tab:sigmaDp}. In Sec.~\ref{sec:hydro}, we applied our formalism to study the hydrodynamics of the critical FS, and we calculated the viscosities and the result is presented in Eqs.\eqref{eq:sigmaq}, \eqref{eq:nuq} and Table.~\ref{tab:Gammaeff}.

    Our computation of the optical conductivity, with results agreeing with the recent perturbative calculation \cite{SLi2023}, also provides a few new perspective to understanding transport:

     First, our formalism extends the Prange-Kadanoff reduction formalism \cite{REPrange1964,HGuo2022a} into a systematic expansion in $\xi/(k_Fv_F)$, and this allows us to perform computation in the NFL regime A.

     Second, a natural picture that emerges from our formalism is that the conductivity is a sum of different conduction channels, which are eigenfunctions of the kinetic operator $L$ in the energy and momentum domain. From our results, we believe the usage of extended Drude formula is not appropriate for momentum conserved systems. The textbook Drude formula \cite{NWAshcroft1976} is derived in the context with momentum relaxation, where the dominant transport lifetime $\tau$ is the momentum relaxation time. In a momentum conserved system, momentum cannot contribute to $\Re\sigma(\omega)$ at the leading order, and we found several mechanisms for $\Re\sigma(\omega)$:
     \begin{enumerate}
       \item In the NFL regime A, the overlap between the current and the momentum has non-analytic dependence on $|\omega|$, and this is reflected as a correction to the Drude peak $\sigma_D$.
       \item For a non-circular FS, the current does not fully overlap with the momentum at zero energy. The remaining part of the current operator is the soft modes of the FS which describe its shape fluctuations. These soft modes lead to a Drude-like contribution  $\sigma_D'$ where the transport lifetime is the inverse eigenvalue of these soft modes.
       \item The current also overlaps with other finite-energy channels and they contribute to the incoherent conductivity $\sigma_i$. This is an explicit example that invalidates the Drude picture because the overlap between current and these incoherent channels vanishes at zero energy. In previous literature there  is a wrong conception that the $\sigma_i\sim T^4\ln T/\omega^2$ incoherent conductivity in the FL regimes C and D can be interpreted as a scattering rate $\gamma\sim T^4\ln T$ via the extended Drude formula, and so it was conjectured to be related to the soft-mode scattering rate which also scales as $T^4$ for a convex, large-angle scattering FL. This interpretation is incorrect because the $\sigma_i$ can appear for any FS geometry where the $T^4$ soft-mode scaling does not always apply. The correct interpretation provided by our calculation is that the $T^2$ factor in $\sigma_i$ is due to the energy-dependent overlap between the current and the conduction channel, and the transport lifetime of that channel is actually $T^2\ln T$ which is the natural expectation from self-energy.
     \end{enumerate}
     Perturbative calculations can in principle obtain the same total conductivity which is the sum of the above mechanisms, but it cannot isolate each individual contributions because perturbative calculations are directly evaluating the inverse of the kinetic operator $L$, so all eigenvectors are mixed together.

     Because the (homogeneous) optical conductivity contains various conduction channels, it is not convenient as a probe for the critical FS physics.  Instead, the non-local conductivity $\sigma(\vec{q})$ measured in the hydrodynamics regime is much cleaner in the sense that it is dominated by the propagation of the FS soft modes. Therefore,  the viscosities we calculated in Eq.\eqref{eq:nuq} and Table.~\ref{tab:Gammaeff} are a more direct signature of the NFL physics.

\begin{acknowledgments} We thank Dmitrii L. Maslov, Andrey V. Chubukov and Alex Levchenko for the inspiring discussions that initiated this work at the KITP program ``Quantum Materials With And Without Quasiparticles'' . KITP is supported in part by the National Science Foundation under Grants No. NSF PHY-1748958 and PHY-2309135. We thank Debanjan Chowdhury, Zhengyan Darius Shi, Hart Goldman, Senthil Todadri, Leonid Levitov, J\"org Schmalian, Aavishkar A. Patel, Ilya Esterlis and Subir Sachdev for helpful discussions. Haoyu Guo is supported by the Bethe-Wilkins-KIC postdoctoral fellowship at Cornell University.
    \end{acknowledgments}

\bibliography{NFL,supp}
\end{document}